\documentclass[oldversion]{aa}
\usepackage{epsfig}
\usepackage{natbib}
\usepackage{lscape}
\usepackage{wasysym}
\bibpunct{(}{)}{;}{a}{}{,}
\begin{document}

\title{A new Neptune-mass planet orbiting HD\,219828\thanks{Based on 
observations collected
at the La Silla Parana Observatory,
ESO (Chile) with the HARPS spectrograph at the 3.6m telescope
(Observing runs 075.C-0332, 076.C-0155, and 77.C-0101).}}


\author{
  C. Melo\inst{1} \and
  N. C. Santos\inst{2,3,4} \and
  W. Gieren\inst{5} \and
  G. Pietrzynski\inst{5} \and
  M. T. Ruiz\inst{6} \and
  S. G. Sousa\inst{2,7,8} \and
  F. Bouchy\inst{9} \and
  C. Lovis\inst{3} \and
  M. Mayor\inst{3} \and
  F. Pepe\inst{3} \and
  D. Queloz\inst{3} \and
  R. da Silva\inst{3} \and
  S. Udry\inst{3}
}

\institute{
    European Southern Observatory, Casilla 19001, Santiago 19, Chile
    \and
    Centro de Astronomia e Astrof{\'\i}sica da Universidade de Lisboa,
    Observat\'orio Astron\'omico de Lisboa, Tapada da Ajuda, 1349-018
    Lisboa, Portugal
    \and
    Observatoire de Gen\`eve, 51 ch.  des Maillettes, CH--1290 Sauverny, Switzerland
    \and
    Centro de Geofisica de \'Evora, Rua Rom\~ao Ramalho 59, 7002-554 \'Evora, Portugal
    \and
    Universidad de Concepcion, Departamento de Fisica, Casilla 160-C, Concepcion, Chile
    \and
    Departamento de Astronomia, Universidad de Chile, Casilla Postal 36D, Santiago, Chile
    \and
    Centro de Astrof{\'\i}sica da Universidade do Porto, Rua das Estrelas, 4150-762 Porto, Portugal
    \and
    Departamento de Matem\'atica Aplicada, Faculdade de Ci\^encias da Universidade do Porto, Portugal
    \and    
    Institut d'Astrophysique de Paris, 98bis Bd. Arago, 75014 Paris, France
}

\date{Received XXX; accepted XXX}

\abstract{Two years ago a new benchmark for the planetary survey was set with the 
discoveries of three extrasolar planets with masses below 20$M_\oplus$. In particular, 
the serendipitous discovery of the 14$M_\oplus$ planet around $\mu$ Ara found with HARPS
with a semi-amplitude of only 4~m\,s$^{-1}$ put in evidence the tremendous potential 
of HARPS for the search of this class of very low-mass planets.
Aiming to discovering new worlds similar to  $\mu$\,Ara\,b, we carried out an intensive 
campaign with HARPS to observe a selected sample of northern stars covering a range of 
metallicity from about solar to twice solar. Two stars in our program were found to 
present radial velocity variations compatible with the presence of a planet-mass companion.
The first of these, HD\,219828, was found to be orbited by a planet with a 
minimum mass of 19.8\,$M_\oplus$ and an orbital period of 3.83 days. It is the 11th
Neptune-mass planet found so far orbiting a solar-type star. The radial velocity 
data clearly show the presence of an additional body to the system, likely of planetary mass. 
The second planet orbits HD\,102195, has a mass of 0.45$M_{Jup}$ and an orbital 
period of 4.11 days. This planet has been already announced by Ge et al. (2006). 
Our data confirm and improve the orbital solution found by these authors.
We also show that the high residuals of the orbital solution are caused by stellar 
activity, and use the bisectors of the HARPS cross-correlation
function to correct the noise introduced by stellar activity. An improved 
orbital solution is obtained after this correction. This kind of analysis 
may be used in the future to correct the radial-velocities for stellar 
activity induced noise.
  }

\authorrunning{Melo et al.}
\maketitle

\section{Introduction}

The serendipitous discovery by \citet{Santos-2004a} of a 14M$_\oplus$ planet  
orbiting $\mu$\,Ara, simultaneously followed by the announcement 
of another planet of similar mass around 55 Cnc \citep{McArthur-2004} and of
a 21M$_\oplus$ mass planet around GJ 436 b \citep{Butler-2004} set a new benchmark 
for planet surveys. Since then, a few more cases have been announced 
\citep[][]{Rivera-2005,Bonfils-2005,Vogt-2005,Udry-2006},
including a system of three Neptune-mass bodies \citep[][]{Lovis-2006}. 
For the first time in the literature, the minimum masses of a planet 
found by Doppler shift techniques were indicated in Earth-masses and not in Jupiter- 
or Saturn-masses. Even more exciting was the fact that based on their orbital characteristics 
and masses, it is strongly suggested that these planets may be rocky or icy in 
nature \citep[][]{Alibert-2006}. These discoveries may have thus unveiled the
first super-earths, and shown that telluric planets may be common in the
solar neighborhood. 

The detection of these kinds of planets is not easy to be achieved. For example, 
the discovery of the 14\,M$_\oplus$ companion of $\mu$\,Ara was only made 
possible thanks to the two 
main factors. First, the intrinsic stability of an instrument like HARPS 
which is able to keep a long-term (several years) radial velocity accuracy 
better than 1 m\,s$^{-1}$. Second, the observational strategy. $\mu$\,Ara was 
observed in a context of a asteroseismology program, where a unique 
target was monitored during the whole night and a large number of spectra was collected with 
exposure times of about a 2 minutes, in order to measure the stellar acoustic 
oscillation modes. Seen from the perspective of a radial velocity planetary search, 
these oscillations are unwanted and are considered as stellar noise. This intrinsic noise
is well illustrated by the case of $\mu$\,Ara: during the nights when the star was 
measured with HARPS over a several hours period, the residuals around the orbital fit
were merely around 0.4\,m\,s$^{-1}$, a value that increased to near 1.5\,m\,s$^{-1}$
for the remaining nights, when the radial-velocity of the star was
an average of only about 15 minute integrations \citep[][]{Santos-2004a}; this value 
would be around 3\,m\,s$^{-1}$ if only one single (short exposure) radial-velocity 
measurement was done. In other words, to be able to achieve the 1\,m\,s$^{-1}$ precision 
one needs to average the radial velocities for a given star over periods of several 
minutes (at least 15-20 minutes), making the searches for very low mass planets 
very time consuming \citep[see the seismological results on $\mu$ Ara by][]{Bouchy-2005c}. 

Surprisingly enough, $\mu$\,Ara and 55\,Cnc were back then the only two stars that were 
(incidentally) measured with enough precision so that such low mass planets could have
been detected. This may imply that very low mass short period planets are very common. 
This result is supported by recent simulations \citep[][]{Ida-2004b,Benz-2006} that 
suggest that very low-mass planets may be more frequent than the previously 
found giant worlds. 

These theoretical results and the discovery of several Neptune-mass planets 
suggested that if solar type stars are correctly monitored, 
the number of detection of such low-mass rocky planets can be very high. 
Having this in mind, we started a project to use the HARPS spectrograph to monitor 
a sample of about 40 northern stars with metallicity ranging from solar to about 
twice solar. 
In the present paper, we present the discovery of a new Neptune-mass planet
orbiting the star \object{HD\,219828}, as well as the confirmation of a Jovian-mass
planet orbiting \object{HD\,102195}. We also find evidence for the existence of a longer
period planet companion to the former star.
The presentation of the whole sample, along with a
detailed study of the behavior of the radial velocities with 
other properties (e.g. spectral type and activity level) is postponed to a future paper.

\section{Observations}

The observations were carried out using the HARPS spectrograph (3.6-m ESO telescope, 
La Silla, Chile), in three different observing runs between May 2005 and July 2006 
(ESO observing runs 075.C-0332, 076.C-0155, and 77.C-0101). Further measurements of the two
stars discussed in this paper were also done in different HARPS GTO runs, in collaboration with 
the Geneva team.

Radial-velocities were obtained directly using the HARPS pipeline. We 
refer to \cite{Pepe-2004} for details on the data reduction. The individual
spectra were also used to derive both the Bisector Inverse Slope (BIS) of the 
HARPS Cross-Correlation Function (CCF), as defined by \cite{Queloz-2001}, 
as well as a measurement of the chromospheric activity index $\log{R'_{\rm HK}}$, 
following the same recipe used by \citet[][]{Santos-2000a} for
CORALIE spectra. Finally, the combined high S/N HARPS spectra were analyzed to
derive stellar atmospheric parameters and iron abundances using the method described 
in \citet[][]{Santos-2004b}.

%
%

\section{A new planet around HD\,219828}

\subsection{Stellar characteristics}

According to the Hipparcos catalog \citep[][]{ESA-1997}, HD\,219828 
is a G0IV star with a parallax $\pi=12.33\pm1.01$ mas, an apparent magnitude 
$m_v=8.04$, and a colour index $B-V = 0.654$. From these parameters,
and taking the bolometric correction from \citet[][]{Flower-1996},
we derive an absolute magnitude $M_v = 3.49$ and a luminosity of 3.34\,$L_\odot$.

The analysis of our combined HARPS spectra, with a total S/N ratio above 500, 
provide $T_{eff} = 5891\pm 18$, $\log g=4.19\pm 0.02$, and $[Fe/H]=0.19\pm 0.02$. 
Along with the above mentioned absolute magnitude, the Geneva Evolutionary 
models \citep[][]{Schaerer-1993a} indicate that HD\,219828 
has a mass of 1.24$M_\odot$ and an age of 5.2\,Gyr. In Table~\ref{table:hd219828_star} we 
summarize the stellar parameters derived for this star.

\begin{table}
\par
\caption{
\label{table:hd219828_star}
Stellar parameters for \object{HD\,219828}. }
\begin{tabular}{lcc}
\hline\hline
\noalign{\smallskip}
Parameter  & Value & Reference \\
\hline
Spectral~type   		& G0IV			& Hipparcos  \\
Parallax~[mas]  		& 12.33$~\pm~$1.01 	& Hipparcos \\
Distance~[pc]   		& 81.1                  & Hipparcos \\
$m_v$           		& 8.04 			& Hipparcos \\
$B-V$  				& 0.654                 & Hipparcos \\
$M_{v}$  			& 3.49 			& -- \\
Luminosity~$[L_{\odot}]$  	& 3.34$\dagger$	        & -- \\
Mass~$[M_{\odot}]$  		& 1.24			& \citet[][]{Schaerer-1993a} \\
$\log{R'_{\rm HK}}$ 		& $-$5.04         	& HARPS \\
$v~\sin{i}$~[km~s$^{-1}$] 	& 2.9$\dagger\dagger$ 	& -- \\
$T_{\rm eff}$~[K]  		& 5891$~\pm~$18 	& -- \\
$\log{g}$  			& 4.18$~\pm~$0.02 	& -- \\
$\xi_{\mathrm{t}}$  		& 1.18$~\pm~$0.02 	& -- \\
${\rm [Fe/H]}$  		& $+$0.19$~\pm~$0.03 	& -- \\
\hline
\noalign{\smallskip}
\end{tabular}
\newline
$\dagger$ Using the bolometric correction of \citet{Flower-1996}
\newline
$\dagger$$\dagger$ From HARPS spectra using a calibration similar to the one presented by \citet{Santos-2002a}
\end{table}

HD\,219828 shows a slightly lower spectroscopic gravity as compared to main-sequence 
dwarf stars of the same temperature. An even lower gravity ($\log g=4.03$) is obtained if 
the isochrone mass is used in conjunction with the stellar parallax \citep[e.g.][]{Santos-2004b}.
This result is compatible with the spectral classification of the star \citep[G0IV --][]{ESA-1997}, and indicates that HD\,219828 may be slightly evolved. 
The stellar radius estimated using the luminosity--temperature--radius relation 
is 1.76$R_\odot$.

\begin{table}[t]
\caption{Radial-velocity measurements of HD\,219828 and HD\,102195.}
\label{table:rv}
\begin{tabular}{ccc}
\hline\hline
JD     &$V_r$ [km\,s$^{-1}$] &$\sigma(V_r)$ [km\,s$^{-1}$]\\
\hline
\multicolumn{3}{l}{HD\,219828}\\
2453509.928056 & -24.03256 & 0.00066\\
2453510.928367 & -24.01790 & 0.00091\\
2453701.542843 & -24.02993 & 0.00041\\
2453702.532514 & -24.01817 & 0.00046\\
2453703.531084 & -24.01824 & 0.00036\\
2453704.530683 & -24.02639 & 0.00040\\
2453705.544423 & -24.02791 & 0.00035\\
2453706.530181 & -24.01890 & 0.00045\\
2453707.529860 & -24.02140 & 0.00059\\
2453708.551863 & -24.03220 & 0.00046\\
2453709.539910 & -24.02877 & 0.00057\\
2453710.539622 & -24.01823 & 0.00049\\
2453930.824911 & -24.05650 & 0.00066\\
2453931.779782 & -24.05201 & 0.00104\\
2453932.776323 & -24.04571 & 0.00065\\
2453933.735140 & -24.04473 & 0.00080\\
2453934.739643 & -24.05830 & 0.00070\\
2453935.739712 & -24.05720 & 0.00069\\
2453936.730288 & -24.04521 & 0.00076\\
2453946.762931 & -24.06075 & 0.00071\\
2453951.833961 & -24.05168 & 0.00074\\
2453975.734459 & -24.05942 & 0.00069\\
\hline
\multicolumn{3}{l}{HD\,102195}\\
2453501.574413 & 2.15457 & 0.00037\\
2453503.580640 & 2.06803 & 0.00133\\
2453504.587954 & 2.15706 & 0.00082\\
2453506.621511 & 2.08172 & 0.00146\\
2453550.535485 & 2.17258 & 0.00075\\
2453551.523110 & 2.11355 & 0.00117\\
2453757.789800 & 2.06038 & 0.00038\\
2453761.827482 & 2.05137 & 0.00036\\
2453765.771759 & 2.07657 & 0.00037\\
2453785.772065 & 2.13058 & 0.00037\\
2453788.796126 & 2.16390 & 0.00033\\
2453927.478651 & 2.08383 & 0.00062\\
2453930.461429 & 2.05648 & 0.00062\\
2453931.470554 & 2.06908 & 0.00067\\
2453932.485026 & 2.17049 & 0.00065\\
2453933.465358 & 2.18322 & 0.00069\\
2453934.470134 & 2.08446 & 0.00062\\
2453935.471600 & 2.06340 & 0.00065\\
2453936.476701 & 2.13495 & 0.00061\\
\hline
\end{tabular}
\end{table}

From the HARPS spectra we derived both a chromospheric activity index ($\log{R'_{\rm HK}}=-5.04$)
and an estimate of the projected rotational velocity of the star ($v~\sin{i}=2.9$ km s$^{-1}$).
From the activity level and the $B-V$ colour
we derive an age of 6.5\,Gyr \citep[][]{Henry-1996} \citep[at least above 2\,Gyr --][]{Pace-2004},
and a rotational period of 26~days \citep[][]{Noyes-1984}. All these values 
suggest that HD\,219828 is an old chromospherically quiet star. According to the Hipparcos
catalog, this star is considered to be constant in photometry, with a scatter of 
only 0.013\,mag (typical for a constant star of its magnitude).

\subsection{HARPS orbital solution}

HD\,219828 was observed 22 times with the HARPS spectrograph, 
between May 2005 to August 2006. Each measurement was done using an exposure
time of 900\,s, in order to average out the stellar oscillation 
noise \citep[e.g][]{Bouchy-2005c}. The complete radial velocity measurements obtained and the errors are presented in Table\,\ref{table:rv}. It is worth noticing that the errors quoted in Table\,\ref{table:rv}, which are used to plot the error bars, refer solely to the instrumental (calibration) and photon-noise error 
share of the total error budget (e.g. activity and/or stellar oscillations are not considered, 
given the difficulty in having a clear estimate of their influence).

\begin{figure}[t]
\resizebox{\hsize}{!}{\includegraphics{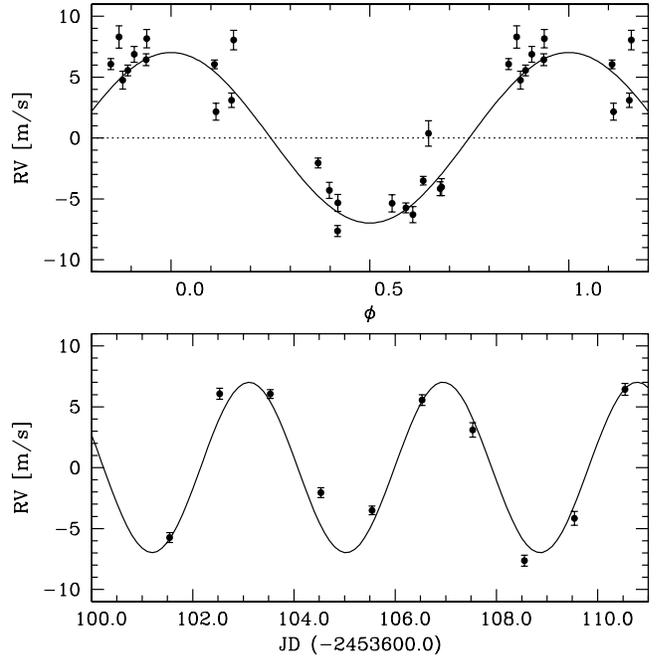}}
\caption{{\it Top}: Phase-folded radial-velocity measurements of HD219828, and
the best Keplerian fit to the data with a period of 3.8-days. In this fit the
long term radial-velocity trend was subtracted. See text for more details.
{\it Bottom}: Radial-velocity measurements of HD219828 as a function of time for one
epoch of measurements.}
\label{fig:hd219828_phase}
\end{figure}

\begin{table}[b]
\caption[]{Elements of the fitted orbit for HD\,219828b.}
\begin{tabular}{lll}
\hline
\hline
\noalign{\smallskip}
$P$             & 3.8335$\pm$0.0013    				  & d\\
$T$             & 2453898.6289$\pm$0.072   			  & d\\
$e$             & 0.0$^\dagger$     				  &  \\
$V_r$           & $-$24.032$\pm$0.001    			  & km\,s$^{-1}$\\
$\omega$        & 0.0$^\dagger$       				  & degr \\ 
$K_1$           & 7.0$\pm$0.5     				  & m\,s$^{-1}$ \\
$f_1(m)$        & $0.1364\cdot10^{-12}$$\pm$$0.3572\cdot10^{-12}$ & M$_{\odot}$\\ 
$\sigma(O-C)$   & 1.7		                      		  & m\,s$^{-1}$  \\    
$N$             & 22                                  		  &  \\
$m_2\,\sin{i}$  & 19.8                                		  & M$_{\oplus}$\\
\noalign{\smallskip}
\hline
\end{tabular}
\\$^\dagger$ Fixed to zero. The obtained eccentricity was consistent with a circular 
orbit according to the \citet[][]{Lucy-1971} test.
\label{table:hd219828_orbit}
\end{table}

A quick analysis of the data revealed the presence of a stable 3.83-day 
period radial-velocity signal. This signal is best fitted using a Keplerian fit 
with an amplitude K of 7.0\,m\,s$^{-1}$, and a non-significant eccentricity 
(see Fig.\,\ref{fig:hd219828_phase}; leaving the eccentricity as a free parameter 
in the orbital fit provides a value of 0.11$\pm$0.09). This corresponds to the 
expected signal induced by the presence of a 19.8 Earth-mass (minimum-mass)
companion to HD\,219828 (Table\,\ref{table:hd219828_orbit}).

\begin{figure}[t]
\resizebox{\hsize}{!}{\includegraphics{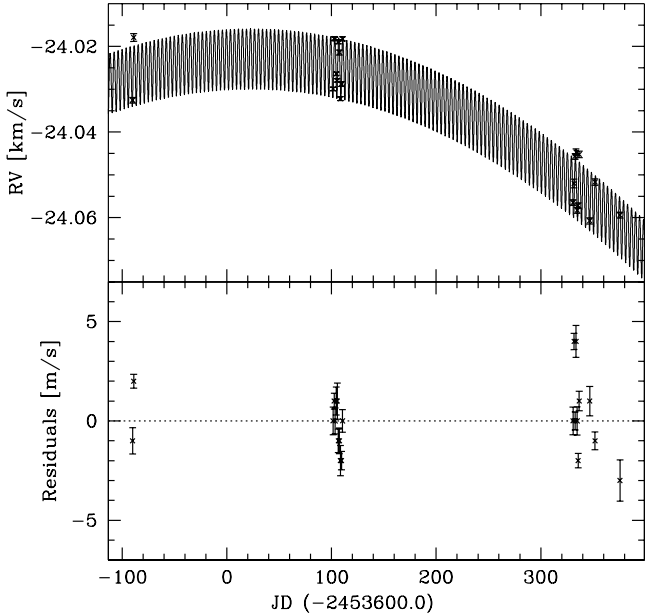}}
\caption{{\it Top}: Radial-velocity measurements of HD219828 as a function of time, and
a fit to the data including a 3.8-day period Keplerian and a long period quadratic trend.
{\it Bottom}: Residuals of the fit.}
\label{fig:hd219828_time}
\end{figure}

Our radial velocity measurements as a function of Julian Date are shown 
in Fig~\ref{fig:hd219828_time}. The plot shows that, in addition to the short 
period signal, there is a long-period trend superimposed. 
We have tried to fit 
a two Keplerian model to the whole radial-velocity set using the {\it stakanof} 
genetic algorithm, recently developed by Tamuz et al. \citep[in preparation; 
see also ][]{Pepe-2006}. Although the 3.83-day period signal was always present 
in the solutions, several similar quality solutions 
were found for the long period signal. These always ranged from $\sim$180 to $\sim$800 days 
in period, corresponding to the presence of a planet in the Jupiter-mass domain ($\sim$0.7\,M$_{\mathrm{Jup}}$).
The best fit (with an rms of only 1.2\,m\,s$^{-1}$) gave a period of 181-days,
eccentricity of 0.3, and amplitude K=21.6\,m\,s$^{-1}$ for the long period orbit.

Given the ambiguity in the result, we prefer at this point to
fit a simple quadratic drift to the residuals of the short period signal. The global
rms of the Keplerian $+$ drift solution is 1.7\,m\,s$^{-1}$.
We note, however, that this long term trend does not perfectly fit
the observed variation; some of the two Keplerian models had better $\chi^2$ than
the adopted preliminary solution. This is also illustrated by the increasing residuals of
the fit as a function of time (Fig.\,\ref{fig:hd219828_time}). These facts explain the 
relatively high rms of the orbital solution found, and may imply that the
orbital parameters (e.g. $e$ and $T$) of the short period planet may change slightly when
the long period solution is better constrained.

\begin{figure}[t]
\resizebox{\hsize}{!}{\includegraphics{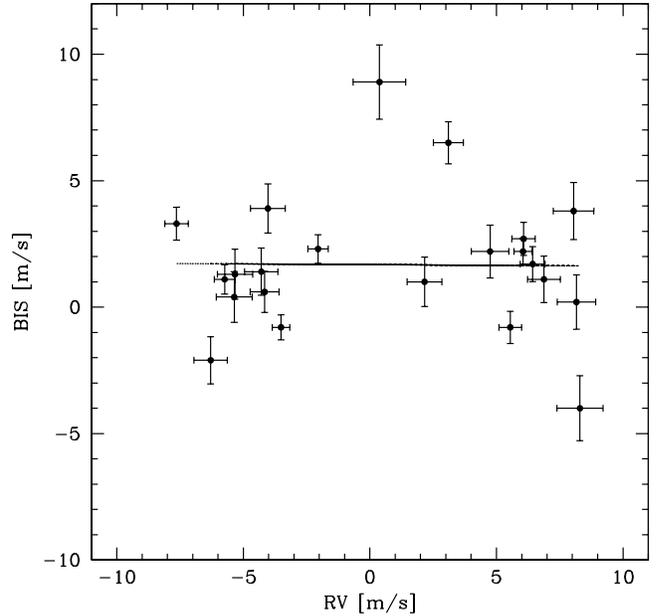}}
\caption{BIS vs. radial-velocity for HD\,219828. The best linear fit to the data is shown,
and has a slope compatible with zero.}
\label{fig:hd219828_bis}
\end{figure}

To understand if the short period radial-velocity signal observed could be due to
the presence of stellar spots \citep[e.g.][]{Saar-1997,Queloz-2001} or stellar
blends \citep[][]{Santos-2002a}, we computed the Bisector Inverse Slope (BIS) of the 
HARPS Cross-Correlation Functions (CCF) of HD\,219828. In Fig.\,\ref{fig:hd219828_bis} 
we plot the resulting BIS as a function of the radial-velocity, after having
subtracted the long period quadratic trend to the data. The result shows that
no correlation exists between the two quantities, suggesting that stellar activity
of stellar blends cannot explain the short period and low amplitude radial-velocity
variation observed.

Together with the very low activity level of the star, we thus conclude that the
3.83-day orbital period observed can be better explained by the presence of
a Neptune-mass planet orbiting HD\,219828.

%
%

\section{The planet around HD\,102195}

\subsection{Stellar characteristics of HD~102195}

According to the Hipparcos catalog \citep[][]{ESA-1997}, 
HD\,102195 is a K0V star with visual magnitude $m_v=8.07$ and colour index $B-V=0.835$.
From the visual magnitude and astrometric distance of 28.98 pc ($\pi=34.51\pm1.16$),
we derive an absolute magnitude $M_v = 5.76$. Considering the bolometric correction taken
from \cite{Flower-1996}, the obtained luminosity is then 0.47\,$L_\odot$. 

Accurate stellar atmospheric parameters for HD\,102195 were derived from the 
analysis of our combined HARPS spectra, with a S/N ratio above 300. Our analysis 
provides $T_{eff}=5291\pm34$K, $\log g=4.45\pm0.04$, and $[Fe/H]=0.05\pm0.05$, 
in excellent agreement with
previous estimates \citep[][]{Ge-2006}. 
This value for the gravity is compatible with the one derived using the temperature,
astrometric distance, and stellar luminosity ($\log g=4.56$).
A stellar mass of 0.87$M_\odot$ is derived from the 
absolute magnitude, effective temperature, and metallicity, after comparison 
with stellar evolution models \citep[][]{Schaller-1992}.
The stellar radius estimated using the luminosity--temperature--radius relation 
is 0.82$R_\odot$. According to the properties summarized in Table~\ref{table:hd102195_star}, 
HD\,102195 is a typical solar metallicity main-sequence dwarf. 

The HARPS cross-correlation function gives an estimate of the projected stellar 
rotational velocity $v\,\sin{i}=2.6$\,km\,s$^{-1}$, a value that is reasonably compatible 
with the one derived by \citep[][]{Ge-2006}. Using HARPS spectra, we further
derive a rather high value for the chromospheric activity index for this 
star ($\log{R'_{\rm HK}}=-4.56$). A similarly high value of $\log{R'_{\rm HK}}=-4.30$ 
was obtained by \citet[][]{Strassmeier-2000}.

Using the derived chromospheric activity index, and the colour of the star, we
used the \citet[][]{Henry-1996} calibration to derive an estimate for the 
stellar age of 1.17\,Gyr, and a rotational period of about 20-days \citep[][]{Noyes-1984},
slightly above the value found using the photometric light curve \citep[$\sim$12-days -- ][]{Ge-2006}.
As an additional age indicator, the Li abundance was computed 
as described in \cite{Israelian-2004}. We found a Li abundance $\log{\epsilon\,(Li)}<0.2$. 
According to \cite{Sestito-2005}, a $T_{eff}\sim5300$K star with its lithium content should have 
an age above 0.5\,Gyr, also compatible with the estimate of \citet[][]{Ge-2006}.

\begin{table}
\par
\caption{Stellar parameters for HD\,102195.}
\begin{tabular}{lcc}
\hline\hline
\noalign{\smallskip}
Parameter  & Value & Reference \\
\hline
Spectral~type   		& K0V 			& Hipparcos \\
Parallax~[mas]  		& 34.51$~\pm~$1.16 	& Hipparcos \\
Distance~[pc]   		& 28.98 		& Hipparcos \\
$m_v$           		& 8.07 			& Hipparcos \\
$B-V$  				& 0.835 		& Hipparcos \\
$M_{v}$  			& 5.76 			& -- \\
Luminosity~$[L_{\odot}]$  	& 0.47$\dagger$ 	& -- \\
Mass~$[M_{\odot}]$  		& 0.87 			& \citet[][]{Schaller-1992} \\
$\log{R'_{\rm HK}}$ 		& $-$4.56         	& HARPS \\
$v~\sin{i}$~[km~s$^{-1}$] 	& 2.6$\dagger\dagger$	& \\
$T_{\rm eff}$~[K]  		& 5291$~\pm~$34 	& \\
$\log{g}$  			& 4.45$~\pm~$0.04 	& \\
$\xi_{\mathrm{t}}$  		& 0.89$~\pm~$0.05 	& \\
${\rm [Fe/H]}$  		& 0.05$~\pm~$0.05 	& \\
\hline
\noalign{\smallskip}
\end{tabular}
\newline
$\dagger$ Using the bolometric correction of \citet{Flower-1996}
\newline
$\dagger$$\dagger$ From HARPS spectra using a calibration similar to the one presented by \citet{Santos-2002a}
\label{table:hd102195_star}
\end{table}

\subsection{The HARPS orbital solution}

The discovery of a giant planet around HD~102195 has been recently announced
by the Exoplanet Tracker team\footnote{http://www.astro.ufl.edu/et/}
using a new concept of instrument combining spectroscopy and interferometry  \citep{Ge-2006}.
Since this star is among our surveyed sample, it is interesting to compare the results obtained by 
\cite{Ge-2006} with those obtained with HARPS.

We observed HD\,102195 four times in May 2005, in the framework of our program 
aimed at finding earth-mass planets with HARPS around a
metallicity biased sample of stars. Based on these first four measurements, 
HD\,102195 was clearly detected to be a radial velocity variable. These first four measurements
did not show any correlation between the bisector and the radial velocities,
suggesting that this star could be indeed a planet host.  

\begin{figure}[t]
\resizebox{\hsize}{!}{\includegraphics{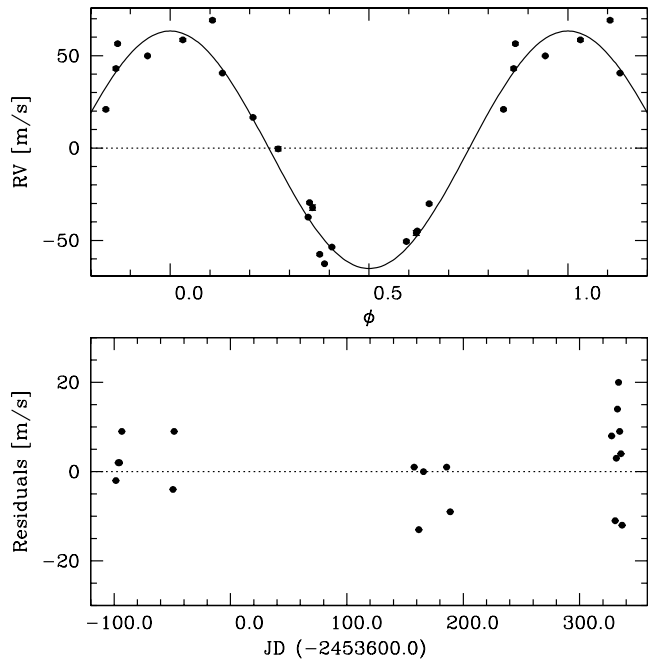}}
\caption{{\it Top}: Phase-folded radial-velocity measurements of HD102195, and
the best Keplerian fit to the data with a period of 4.1-days.
{\it Bottom}: Residuals of the Keplerian fit as a function of time.}
\label{fig:hd102195_phase}
\end{figure}

\begin{figure}[t]
\resizebox{\hsize}{!}{\includegraphics{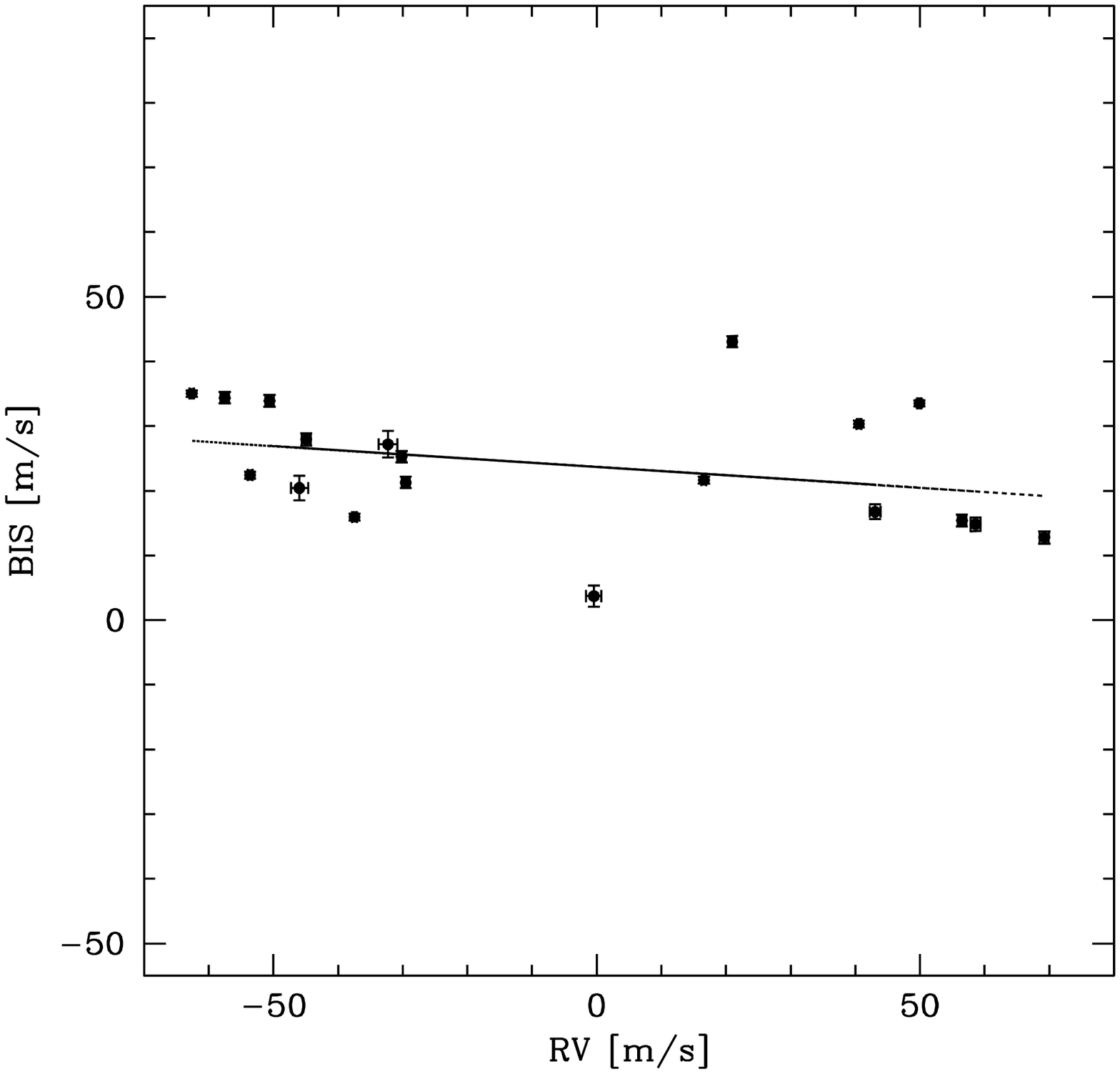}}
\caption{BIS vs. radial-velocity for HD\,102195. The best linear fit to the data is shown,
and has a slope compatible with zero.}
\label{fig:hd102195_bis}
\end{figure}

\begin{figure}[t]
\resizebox{\hsize}{!}{\includegraphics{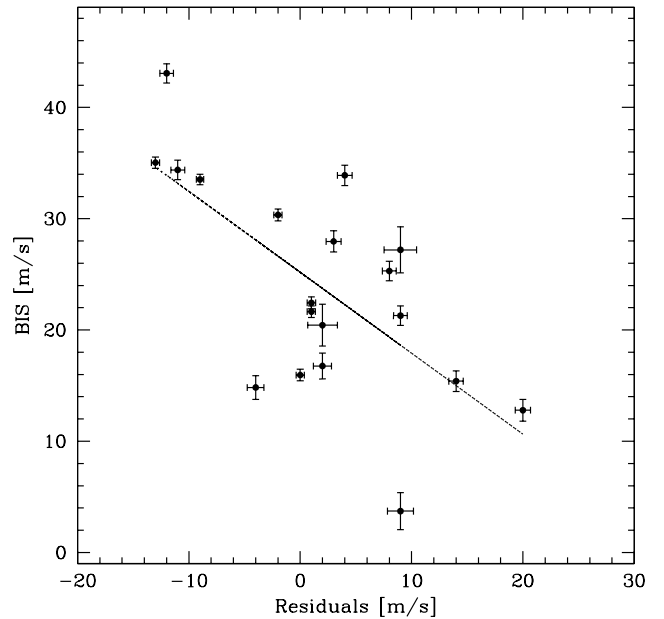}}
\caption{BIS as a function of the residuals of the 4.1-day Keplerian radial-velocity fit for 
HD\,102195. The data is clearly correlated. The best linear fit has a slope of $-0.73$.}
\label{fig:hd102195_ocft_bis}
\end{figure}

The HARPS GTO team was contacted and kindly accepted to monitor HD\,102195. In total, 
and counting also the data points obtained in our subsequent observing runs, 19 
radial-velocity points were added between June 2005 and February 2006 in order to 
derive a good orbital solution. In the meanwhile, the star was announced as a 
planet host by \cite{Ge-2006}.
In Table~\ref{table:rv} we show the radial velocity measurements collected 
for HD\,102195 between May 2005 and February 2006 covering
a time span of about 300 days.

\begin{table*}[t!]
\caption[]{Elements of the fitted orbit for HD\,102195b, using the original and the corrected radial-velocity 
measurements. See text for more details.}
\begin{tabular}{llll}
\hline
\hline
\noalign{\smallskip}
                & \multicolumn{1}{l}{Original velocities} & \multicolumn{1}{l}{Corrected velocities} \\
\hline
$P$            & 4.114225$\pm$0.000855		  & 4.113775$\pm$0.000557 		& d\\
$T$            & 2453896.00$\pm$0.04 		  & 2453895.96$\pm$0.03 		& d\\
$e$            & 0.0$\dagger$			  & 0.0$\dagger$			&  \\
$V_r$          & 2.113$\pm$0.002		  & 2.132$\pm$0.001  			& km\,s$^{-1}$\\
$\omega$       & 0.0$\dagger$ 			  & 0.0$\dagger$			& degr \\ 
$K_1$          & 64$\pm$3			  & 63$\pm$2 				& m\,s$^{-1}$ \\
$f_1(m)$       & $1.135\cdot10^{-10}$$\pm$$0.169\cdot10^{-10}$ & $1.070\cdot10^{-10}$$\pm$$0.103\cdot10^{-10}$ & M$_{\odot}$\\ 
$\sigma(O-C)$  & 9.4		        	  & 6.1  				& m\,s$^{-1}$  \\       
$N$            & 19		        	  & 19   				&  \\
$m_2\,\sin{i}$ & 0.46		        	  & 0.45 				& M$_{Jup}$\\
\noalign{\smallskip}
\hline
\end{tabular}
\\$^\dagger$ Fixed to zero. The obtained eccentricity was consistent with a circular 
orbit according to the \citet[][]{Lucy-1971} test.
\label{table:hd102195_orbit}
\end{table*}

Using the radial velocity data shown in Table~\ref{table:rv} an orbital 
solution was found (Table~\ref{table:hd102195_orbit}). The best fit Keplerian orbit
gives a period of 4.11-days, an amplitude K of 64\,m\,s$^{-1}$, and a non-significant
eccentricity. These values are similar to the ones derived by \citet[][]{Ge-2006}. 
The phase-folded radial velocity plot of the measurements is shown in Fig.~\ref{fig:hd102195_phase}.

Despite the high activity level of the star, an analysis of the bisector
of the HARPS cross-correlation functions shows no clear correlation with
the radial-velocities (Fig.\,\ref{fig:hd102195_bis}). This strongly suggests
that the radial-velocity signal is not being produced by stellar spots or
stellar blends, and supports the planetary explanation.

Using the best-fit orbital solution and the mass of HD\,102195, we derive 
for the companion a minimum mass 0.46\,M$_{Jup}$, in close agreement with the 
0.49M$_{Jup}$ found by \cite{Ge-2006}.

The stellar radius together with the rotational period mentioned above provide a value
for the stellar rotational velocity of $\sim$3.5\,km\,s$^{-1}$. Comparing this with the
estimated value for the projected rotational velocity ($v\,\sin{i}=2.6$\,km\,s$^{-1}$) 
we get $\sin{i}=$ 0.74, corresponding to an inclination of 47\,degrees. If the orbital
axis of the planetary orbit is aligned with the stellar rotation 
axis \citep[e.g.][]{Queloz-2000b}, such an inclination
could explain why no transit of HD\,102195\,b was detected by \citet[][]{Ge-2006}.
The real mass of planet would then be 0.62\,\,M$_{Jup}$.


\subsection{Correcting the radial velocities for the stellar noise}
{

%
%
%

In the case of HD~102195, the residuals (O-C) given in Table~\ref{table:hd102195_orbit} of 9.4 m s$^{-1}$
are much higher than the errors on radial velocity quoted in Table~\ref{table:rv}.
This extra-noise is likely due to stellar activity (see previous section).
\citet[][]{Ge-2006} found that this star has a photometric variation with a period
of $\sim$12 days, and an amplitude of $\sim$0.015 magnitudes.

The signature of its high activity level is also seen on the bisector analysis done
in this paper. Figure \,\ref{fig:hd102195_ocft_bis} shows
that the residuals of the best Keplerian orbit (previous section)
do correlate with the bisector measurements in a clear indication that
the high-activity level of this star is responsible for the observed high 
residuals \citep[e.g.][]{Saar-1997,Santos-2000a}.

According to Fig.~\ref{fig:hd102195_ocft_bis}, the relation between the residual
radial-velocities and the BIS measurements can be approximated by
a linear function $a \times BIS + b$ whose coefficients were determined by a
least-squares fit. This function can be subtracted from the radial velocity data points
in order to correct for the stellar activity. 

%

Using those corrected velocities, a new orbital solution
was found. The results, also presented in Table\,\ref{table:hd102195_orbit},
show a clear improvement. Not only the rms around the orbital fit
was reduced from 9.4 to 6.1\,m\,s$^{-1}$ (equivalent to subtracting quadratically 
a noise of 7\,m\,s$^{-1}$), but also the estimated errors in
all orbital parameters decreased by almost a factor of two. The resulting 
mass for HD\,102195\,b also slightly decreased (0.45\,M$_{Jup}$) due to the small 
decrease in the amplitude of the orbital solution.

This result shows that the study of the BIS
can be used to correct, at least to some extent, the noise on the radial-velocities 
introduced by intrinsic stellar activity. If the BIS vs. RV correlation can be calibrated (observationally or theoretically) as a function of the stellar
  properties, we could think that in the future an externally determined BIS vs. RV function 
  can be used to correct
  the radial velocity data. This will be extremely important in reaching accuracy of the other of
  the cm\,s$^{-1}$ as expected in future instruments like CODEX \citep{Pasquini-2005}.
}

\section{Concluding remarks}

In this paper, we present the discovery of a new 20\,M$_\oplus$ (minimum-mass) short period
planet, orbiting the sun-like star HD\,219828. The presence of a second longer period
companion to the system, likely a Jovian planet, is also clear from our data. 
Unfortunately, the bad phase coverage of the radial-velocity points does not permit 
to settle the orbital parameters of this latter companion. Finally, we confirm 
the existence of a $m_2\,\sin{i}$=0.45\,M$_{Jup}$ planet orbiting HD\,102195, 
previously announced by another team. 

The planet around HD\,219828 is the 11th found with masses similar to the mass of Neptune.
As the number of these systems increases, new statistical analysis will be
possible. In this sense, \citet[][]{Udry-2006} have discussed the 
possibility that the well known strong correlation between the presence of planets and
the stellar metallicity that exists for stars hosting giant 
planets \citep[e.g.][]{Gonzalez-1997,Santos-2001,Santos-2004b,Fischer-2005}
does not seem to be present for their lower mass counterparts.
For very-low mass companions, the metallicity distribution could rather
be flat, something that may be explained by recent models of
planet formation \citep[][]{Ida-2004b,Benz-2006}.
Although the metallicity of HD\,219828 is rather high, we must be careful when
analyzing the numbers, since the planet-search project described in this paper
concentrated its efforts in studying a sample of metal-rich stars only. 

\begin{figure}[t]
\resizebox{\hsize}{!}{\includegraphics{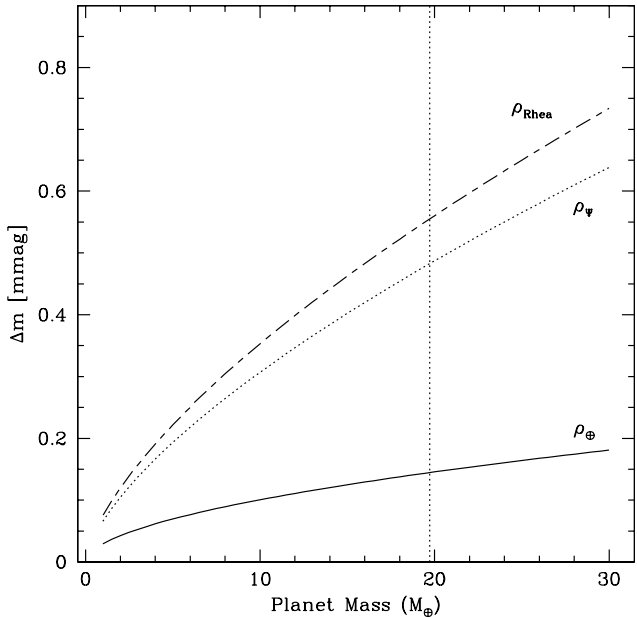}}
\caption{Transit depth expected for three different planet models as a function 
of the mass. The lower curve represents the model of \citet[][]{Valencia-2006},
and the two upper curves the scaled radii considering the mean density of
Neptune and the icy Saturn satellite Rhea. The vertical dotted line represents
the location of HD\,219828\,b.}
\label{fig:depth}
\end{figure}

An important question triggered by the previous discoveries is whether these small planets 
are constituted by rocks, ice or a mixture of both. A more solid assessment of the 
nature of HD\,219828\,b can only be given if the information 
about its mass and radius is available. The detection of a transiting
signature would give us information about the radius and mean density
of the planet, thus opening the possibility to unveil its composition, like is
the case for the giant planets found to transit their host stars \citep[e.g.][]{Konacki-2003,Pont-2004}.

Although the short period of the orbit implies a reasonable value for the transit
probability ($\sim$15\%), the measure of a transit of HD\,219828\,b is not an easy task. 
Using the models presented by \citet[][]{Valencia-2006}, considering that this $m_2\,\sin{i}=19.8\,M_\oplus$ planet could have an Earth-like composition, we 
derive its radius to be $\sim$2.2 Earth radii. 
Although this value must be taken as a lower limit \citep[both due to the lower limit of the mass, and to the fact that it may have an extended atmosphere --][]{Lovis-2006}, observing the 
transit of such a low-mass planet is likely impossible with current ground-based instrumentation
 (see Fig.\,\ref{fig:depth}). 

For the case of HD\,102195, we have also tested, for the first time, the use of the bisector for the 
cross-correlation function to correct the radial-velocity measurements for the noise 
introduced by stellar activity. The results show that a considerable 
improvement can be found. Although the technique may probably be improved,
this conclusion constitutes a first step towards the subtraction of the radial-velocity
noise produced by stellar activity phenomena.

\begin{acknowledgements} 
Support from Funda\c{c}\~ao para a Ci\^encia e a Tecnologia 
(Portugal) to N.C.S. and S.G.S. in the form of fellowships (references
SFRH/BPD/8116/2002 and SFRH/BD/17952/2004) and a grant (reference POCI/CTE-AST/56453/2004) is
gratefully acknowledged. WG, GP and MTR were supported by the Chilean FONDAP Center 
of Astrophysics 15010003.
\end{acknowledgements}

\bibliographystyle{aa}

\end{document}